\documentclass[conference]{IEEEtran}
\IEEEoverridecommandlockouts
\usepackage{cite}
\usepackage{amsmath,amssymb,amsfonts}
\usepackage{algorithmic}
\usepackage{graphicx}
\usepackage[export]{adjustbox}
\usepackage{textcomp}
\usepackage{xcolor}
\usepackage[colorinlistoftodos]{todonotes}
\def\BibTeX{{\rm B\kern-.05em{\sc i\kern-.025em b}\kern-.08em
    T\kern-.1667em\lower.7ex\hbox{E}\kern-.125emX}}
\definecolor{lightblue}{RGB}{128, 130, 202}
\definecolor{darkyellow}{RGB}{223,223,0}

\usepackage{framed}
\usepackage{changepage}
\usepackage{hyperref}

\newcommand{\interviewquote}[2]{
 \def\FrameCommand{%
    \hspace{0pt}%
    {\color{cyan} \vrule width 2pt}
    \colorbox{white}
  }%
  \MakeFramed{\advance\hsize-\width\FrameRestore}%
  \noindent
  \begin{adjustwidth}{}{1pt}
  {\small``\textit{#1}'' - {#2}}\end{adjustwidth}\endMakeFramed%
}

\newcommand{\cristysbox}[1]{\vspace{0.3em}\setlength{\fboxsep}{0.015\linewidth}\noindent\fbox{\parbox{0.96\linewidth}{\vspace{0.1em}#1}} \vspace{0.3em}}

\begin{document}

\title{
Qualifying and Quantifying the Benefits of Mindfulness Practices for IT Workers
}

\author{
\IEEEauthorblockN{\textsuperscript{}Cristina Martinez Montes, Fredrik Sjögren, Adam Klevfors}
\IEEEauthorblockA{\textit{Department of Computer} \\ 
\textit{Science and Engineering} \\
\textit{University of Chalmers$|$Gothenburg}\\
Gothenburg, Sweden \\
montesc@chalmers.se, gussjofrk$|$gusklevad@student.gu.se}
\and
\IEEEauthorblockN{Birgit Penzenstadler}
\IEEEauthorblockA{\textit{Department of Computer} \\ \textit{Science and Engineering} \\
\textit{University of Chalmers$|$Gothenburg}\\
Gothenburg, Sweden \\
\textit{Lappeenranta University of Technology}\\
Lappeenranta, Finland\\
birgitp@chalmers.se}
}


\maketitle

\begin{abstract}
The well-being and productivity of IT workers are crucial for both individual success and the overall prosperity of the organisations they serve. This study proposes mindfulness to alleviate stress and improve mental well-being for IT workers. During an 8-week program, IT workers learn about mindfulness, coupled with breathing practices. This study investigates the potential effects of these practices by analysing participants' reflections through thematic analysis and daily well-being ratings. The analysis showcased an increase in mental well-being and perceived productivity. It also indicated a change in the participants' perception, which showed increased self-awareness. The study recommends continuing the program in the industry to see its impact on work outputs.
\end{abstract}

\begin{IEEEkeywords}
software engineering, mindfulness, mental well-being, breathing practises, stress
\end{IEEEkeywords}

\section{Introduction}
The field of software engineering is known for its fast pace and constant changes, which results in a lot of occupational stress~\cite{amin_software_2011}. `Stress' is defined by the World Health Organisation as a state of worry or mental tension caused by a difficult situation~\cite{who_stress}.  Occupational stress can harm software workers, and have a lasting impact on their mental well-being~\cite{ostberg_methodology_2020}. In the context of Global Software settings, stress has been identified as a factor affecting knowledge sharing and communication among software professionals~\cite{amin_software_2011}. According to Akula et al.~\cite{akula_impact_2008}, work-related stress caused by overtime has a negative impact on software quality. The authors propose that there is a relationship between overtime data and defect count, showing that overtime-driven stress leads to an increase in defect count. They also claim that stress reduces productivity and clarity of mind, resulting in low-quality output.  Furthermore, the stress also has long-term consequences on the health of software developers~\cite{ostberg_methodology_2020}, and their happiness~\cite{graziotin_what_2018}. Another study has shown that stress can also lead to decreased sleep quality and/or deregulation~\cite{martire2020stress}, and sleep deprivation in turn affects a software developer's ability to perform at work~\cite{fucci_need_2020}.

Research within software engineering has, until recently, largely ignored the importance of human factors, such as stress, and puts focus more on technical aspects~\cite{ostberg_methodology_2020}. Yet, humans are still the driving force behind the software. The effects of occupational stress become a problem for both employees and companies when the wanted software quality can not be achieved. 

A lot of studies have showcased the consequences of the stress that the software industry creates, but solutions for it are few and far between. A family of experiments was performed on IT students who used mindfulness techniques over a period of time and positively evaluated on the ability to conceptualize modeling skills~\cite{bernardez2020effects}. 
There have not been enough studies regarding mindfulness as a means for decreasing stress in the software industry, but the few studies conducted have indicated positive results~\cite{bernardez2023empirical}.

Mindfulness in its most simple form is defined by Sim{\'o}n as “the universal and basic human ability to be aware of the contents of the mind moment-by-moment”~\cite{simon_aware_2013}. 
Outside the field of software engineering, Schure et al. report mindfulness and yoga to have beneficial psychological effects on people, such as an increase in their mental clarity, organization, awareness and acceptance of emotions and personal issues~\cite{schure_mindbody_2008}. Furthermore, Poulin reported empirical results of mindfulness training having a positive impact on labour-related environments, especially in stressful ones~\cite{ma_mindfulness_2008}.
Mindfulness has been shown to have a positive impact on people's working lives, but the evidence for its effect within the field of software is still uncertain. 

\paragraph*{Research Objective} The purpose of this study is to explore if the specific breathing practice incorporated in our intervention, based on Yoga pranayama\footnote{energetic breathing exercises, for details on pranayama see~\cite{kupershmidt_definition_2019}}, can positively impact the stress levels and overall mental well-being of IT workers. 

By performing this study, the result might lead to more information about the effects of breathing practices in the IT workspace. Primarily software engineers and other IT workers would benefit for their personal well-being. That said, reducing the effects of stress does not only provide a better working environment but can also reduce errors made by developers. Reducing errors also reduces costs within projects since fixing errors is more costly compared to practices to reduce them. By helping developers to recover better from stress and consecutively to potentially perform better, both the individual and the company benefit in the long run. 

The research questions explored are the following:

\begin{itemize}\itemsep0pt
    \item RQ1: How does the intervention lead to a change in the participant’s \textbf{well-being}?
    \item RQ2: How does the intervention lead to a change in the participant’s \textbf{perceived productivity}?
    \item RQ3: How did the participant’s overall daily \textbf{perception of life} progress over time?
\end{itemize}

The remainder of this paper is structured as follows: Section~\ref{sec:rw} describes the related work, Section~\ref{sec:rm} introduces the research method, Section~\ref{sec:r} explores the results, Section~\ref{sec:d} discusses them, and Section~\ref{sec:c} concludes the paper with a summary and an outlook.

\section{Related Work}\label{sec:rw}

In this section we introduce the related work, starting with stress in the IT world and stress at the workplace, and then moving to interventions for well-being within IT as well as methodological recommendations.

\paragraph*{Stress in the IT world}
Alkubaisi et al. find that work-related stressors are present in everyday work-life and that they negatively influence the performance of employees~\cite{alkubaisi2015can}. For the IT world which is described as a market full of ever-increasing market pressure, long hours of work, tight deadlines, among other things, stress is ``a given'' according to Maudgalya et al.~\cite{maudgalya_workplace_2006}. The authors continue to describe that stress often leads to poor performance and companies being understaffed, meaning there is more overtime to be done by those still working~\cite{maudgalya_workplace_2006}. \newline
The stress and burnout can create working dissatisfaction in the long run for the employee according to Ostberg et al.~\cite{ostberg_methodology_2020}. Sleep deprivation has also been shown as a consequence of stress, and Fucci et al. show that software developers with sleep deprivation performed worse than developers with a full night sleep~\cite{fucci_need_2020}.\newline
Amin et al.~\cite{amin_software_2011} conclude that the globalization of the software industry has increased pressure on software engineers. The pressure can in turn lead to stress for engineers. Further, considering that there are constant tremendous advances in the field of IT, technostress\footnote{Technostress is defined as ``stress experienced by end users in organizations as a result of their use of ICT''~\cite{ragu2008consequences}} is also there to be considered~\cite{bondanini2020technostress}. This highlights the mental well-being of workers within the software industry as an important variable to consider.\newline
Most recently, Wong et al. report from 14 interviews with software engineers to examine mental well-being at work, strategies for managing mental well-being, challenges in using these strategies, and recommendations~\cite{wong2023mental}.

\paragraph*{Impacts of Stress at Work}
As previously mentioned, stress and bad mental well-being lead to worsened working environments for IT workers, which lead to worse performance at work~\cite{lavallee_why_2015}. Moreoever, Lavallé et al.~\cite{lavallee_why_2015} observed IT workers within a telecommunications company for 10 months and saw the decisions made under the pressure of some organizational factors negatively affected software quality. \newline
Aligned, Akula et al.~\cite{akula_impact_2008} confirm that stress in the software industry fosters inadequate results and poor performance by evidencing that stress is correlated to poor performance (1hr and up indirect loss of productivity). Furthermore, the study also implies economic casualties. The most costly part of a project is repairing defects. It is more cost-efficient to prevent defects compared to fixing them. Consequently, the authors believe that increasing stress resilience in IT workers may  affect software quality positively and have a positive economic impact. 

\paragraph*{Well-being Interventions in IT}
Bérnardez et al.~\cite{bernardez_controlled_2014} conducted a quasi-experiment at the University of Sevilla with two groups of software engineering students, one got to practice mindfulness practices and the other went on to practice public speaking. From later conceptual modeling exercises, those who practiced mindfulness scored better on the test than those who practiced public speaking, replicated with a larger group in \cite{bernardez_experimental_2018}. Thus, mindfulness can be confirmed as a practice to increase efficiency within the field of software engineering~\cite{bernardez_experimental_2018}.   
They report their summarized experiments in~\cite{bernardez2020effects}.
More recently, they collected empirical evidence on the positive effect of the practice of mindfulness on a sample of 56 helpdesk employees working for a consulting and information technology company~\cite{bernardez2023empirical}. Furthermore, in their 5th and most recent experiment, they found significant differences in both techno stress and attention awareness between employees of a software company that practice mindfulness and the ones that do not~\cite{bernardez2023impact}. \newline
Last but not least, two of the authors of the paper at hand recently carried out a pilot study intending to compare different well-being intervention modalities, namely yoga poses, breathwork, meditation, and nature walks~\cite{montes2023piloting}.

\paragraph*{Methodology for mental well-being within IT}
Seeing the effect of previously mentioned issues, Ostberg et al.\cite{ostberg_methodology_2020} aspires to provide a methodology to help research regarding stress and mental well-being in the software industry. The proposed methodology was implemented in two studies, where the goal was to analyze metrics such as stress reduction effects, cognitive load, mood improvement, etc. The article provides a good baseline of methodological instruments for research about human factors within IT.

\section{Research Methodology}\label{sec:rm}

\paragraph*{The Rise2Flow program}
Rise2Flow is an intervention centered around a breathing technique to potentially increase participants' well-being, self-efficacy, and perceived productivity~\cite{penzenstadler2022take}. The program does this in a  weekly live session of 90 minutes where the main part of 60 minutes of the time is spent on a specific breathing practice that imitates a physiological sigh and is therefore very relaxing for the nervous system. 
Before that, we provide some introduction and speak about a topic within self-development. The topics covered were: Time Management, Physical Energy, Presence, Recovery, Confidence, Decisions, Listening and Feedback. Each topic comes with success tools (tips) that participants are supposed to use in their daily lives.

\paragraph*{Delimitation} We have previously published partial results of this study in an article that focuses mainly on the quantitative results and only gives a brief preview on qualitative insights~\cite{penzenstadler2022take}. In contrast, the paper at hand goes deep on the qualitative insights from the second time we ran the program (``Rise2Flow2''), especially analyzing the diary entries collected over the entire intervention period.

\emph{Recruitment.}
The participants in this study were recruited by 
personal network and alumni and colleagues from around the globe who were asked to pass on the invite to the study in their networks. Lastly, the study was posted on mailing lists (ICT for Sustainability, LIMITS), Facebook, Instagram, Twitter, LinkedIn, and two large Slack work spaces. The target participants were people within IT, that spend at least 70\% of their working time in front of the computer. Expected participants had at one or several roles of software developers, IT practitioners, IT researchers, IT consultants, IT faculty, and students. The intervention started with 101 registered participants and finished with 33. At least 65 participants wrote a journal entry.

\emph{Data collection.}
The data was collected through online surveys (journals) done daily by each participant. The daily journal is not mandatory, so each participant's data may differ. 
The journal's prompt  was composed of three questions, structured as follows:
\begin{itemize}
    \item Q1: Which well-being exercise did you do today (if any)? Select all that apply: Breathing practice, yoga postures, meditation, nature time, and others.
    \item Q2: How was your day? Select a rating from “Really bad” (1) to “Really great” (10)
    \item Q3: Please write about 100 words: What stood out to you today? What caught your attention? What makes you reflect?
\end{itemize}
The journal, as every other data collection method in this study, prompts the participants to use their anonymous alias for each entry. Their aliases were then combined with demographic data collected about the participants at the beginning of the study.\\
The demographic data consists of age, gender, country of residence, living situation and occupation.

\emph{Data analysis.} \label{Analysis} The data for this study was anonymous, meaning the authors only had access to aliases for each participant and the previously described demographic data. To further ensure confidentiality for the participants, their aliases have been altered.\\
\hspace*{0.4cm}The demographic data served to provide more context to the participants' experiences. This leads to a more in-depth discussion regarding different participants' journal entries.

\subsection{Quantitative}

Using the averages received from the daily journal question, ``How was your day?", the authors proceeded to create visual graphs to display the results acquired, see Fig.~\ref{fig:dailyAverage}.  \\

\subsubsection*{\textit{How does the intervention lead to a change in the participant’s well-being?}} 
\begin{figure}[ht]
\centering
    \includegraphics[width=3cm]{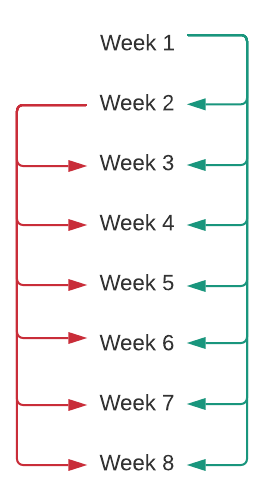}
    \caption{Graphic of the first two iterations of the algorithm were the authors compares all weeks to each other, green arrows were the iterations for week 1 and red arrows were for week 2.}
    \label{fig:stepsAlgorithm}
\end{figure}

We made calculations of the one through ten ratings from the question; ``\textit{How was your day?}" in the daily journal entry. By calculating averages from each week of the eight-week-long study the authors could compare the average well-being week to week. From there the authors could see which week had the highest, as well as the lowest average amongst the weeks. To further see correlations and see between which weeks the average was the most significant, the authors compared each week to another week. For example, starting with week 1. Week 1's average got compared to week 2's average, followed by week 1's average compared to week 3's average, and so on until all weeks had been compared with one another (see Fig.
~\ref{fig:stepsAlgorithm}).

\subsection{Qualitative} \label{Quali}
\begin{figure}[htp!]
\centering
    \includegraphics[width=8cm]{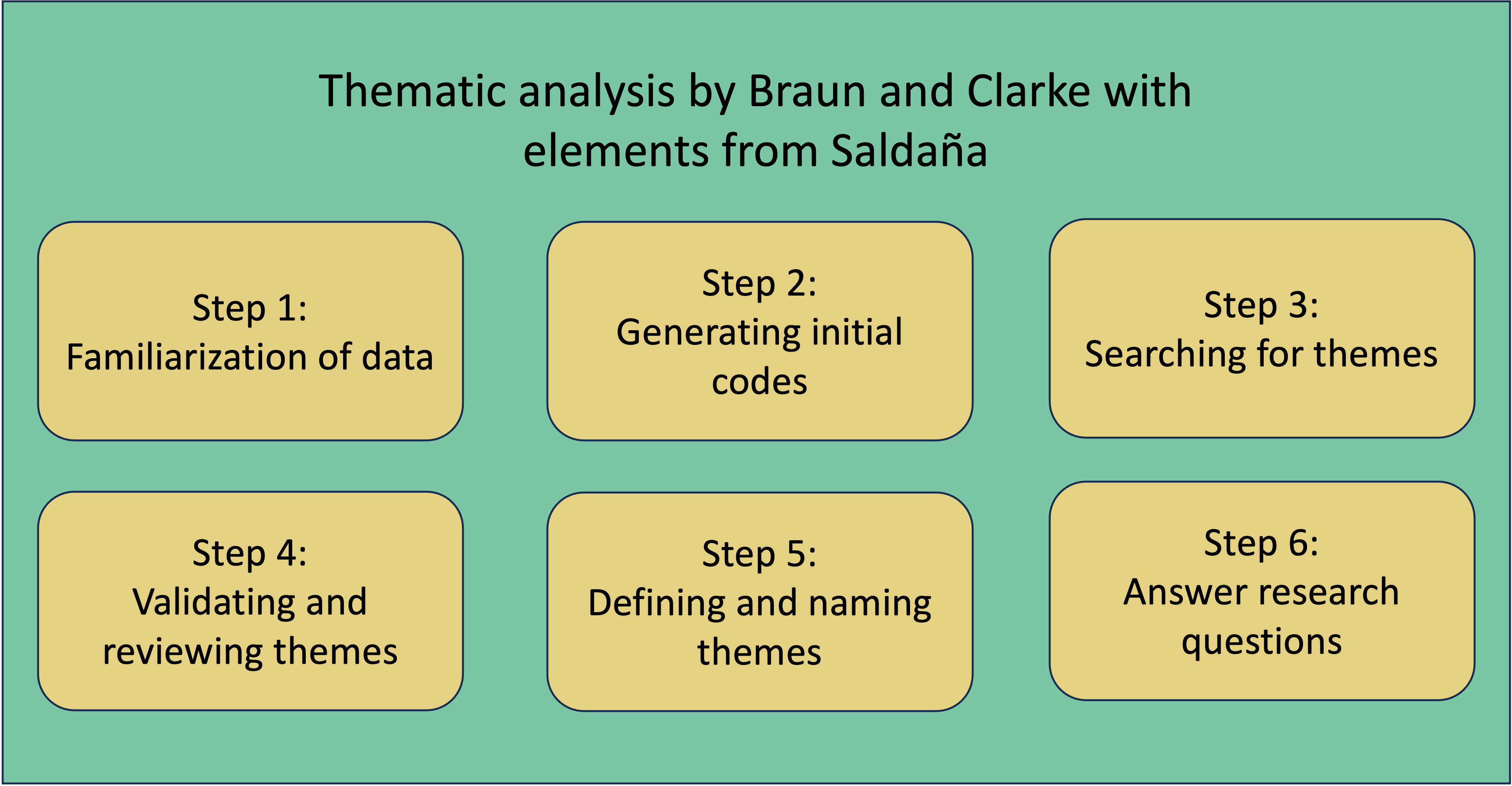}
    \caption{Graphic of the 6 phases of thematic analysis}
    \label{fig:stepsCoding}
\end{figure}

For the qualitative part of the study, the data has been analysed via thematic analysis. The method for this was based on a 6-phase approach by Braun and Clarke~\cite{braun2012thematic}. The stages are summarised in Fig.~\ref{fig:stepsCoding}.

\emph{Step 1) Familiarising Yourself
With the Data}
The two coders familiarised themselves with the data; this was done by reading and rereading all the data. While doing this, both coders started writing down notes on themes found. The text was read with a few questions in mind~\cite{braun2012thematic}: 
"How does this participant make sense of their experiences?", "What assumptions do they make in interpreting their experience?". 
The goal of the first phase was to become familiar with the data set and to begin noticing relevant patterns.\\
\hspace*{0.4cm}\emph{Step 2) Generating Initial Codes}
The initial coding serves as a set of shorthand commands that are mutually agreed upon and understood by both coders. It is important to note that the likelihood of the code being more descriptive than interpretative is acknowledged~\cite{braun2012thematic}. 
The coders opted to use Google Sheets as working in real-time and simultaneously with each other was prioritised over learning new software. When all data had been analysed and coded once, the process was repeated until all data relevant to the study had been collated.\\
\hspace*{0.4cm}\emph{Step 3) Searching for Themes}
By grouping codes into higher-level topics that connected and captured significant data, codes became themes.  The themes together now provided an overview of the data set as a whole.\\
\hspace*{0.4cm}\emph{Step 4) Reviewing Potential Themes}
The fourth phase was quality control of the themes created. The themes were reviewed based on a few questions regarding relevance, inclusivity for each theme, and coherence. While reviewing, a few sub-themes were identified. After the refinement of initial themes, overarching themes and sub-themes were defined.
The phase is described as extra important, especially with big data sets~\cite{braun2012thematic}. The phase served as an opportunity to tweak, delete, or combine previously established themes.\\
\hspace*{0.4cm}\emph{Step 5) Defining and Naming Themes}
During this quality control, the coders had to motivate the naming and defining of themes. They summed up each theme in a few sentences. Each theme had a specific scope, focus, and purpose. Themes should generally have a singular focus~\cite{braun2012thematic}, but not overlap with other themes. The phase included picking extracts for each theme to be analysed. \\
\hspace*{0.4cm}\emph{Step 6) Answering the research questions}
All phases completed were then compiled to answer the research questions. The themes were presented in accordance with which research question it was relevant for. The themes were introduced with an overview of their content and examples to provide the reader with an understanding of each specific theme.

\section{Results and Findings}\label{sec:r}

This section presents the results of the thematic analysis and the analysis of the participants' daily ratings. Table~\ref{tab:themes} displays each theme and corresponding sub-themes. Each theme will be explored further, accompanied by examples and definitions.

\subsection{Daily ratings}

\begin{figure}[htp!]
    \title{Average daily ratings by week}
    \includegraphics[width=9cm]{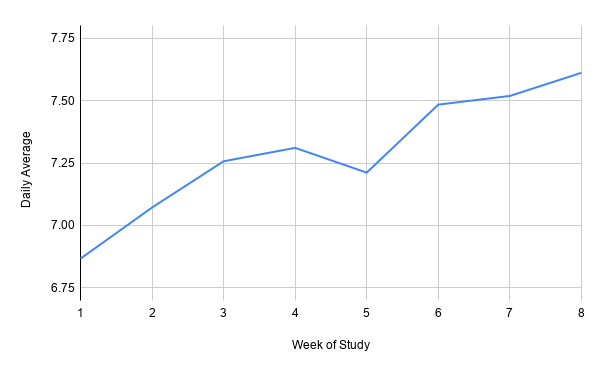}
    \caption{Graph of average daily ratings by week}
    \label{fig:dailyAverage}
\end{figure}

Figure \ref{fig:dailyAverage} describes the participant's answer to Q2 in the daily journal:\textit{ How was your day? Select a rating from “Really bad” (1) to “really great” (10).} Overall, the daily journal consisted of 610 entries. 65 different participants have left at least one journal entry. \\
\hspace*{0.4cm}As seen in Fig.~\ref{fig:dailyAverage}, the daily average of the participants saw an overall increase over the eight weeks. In week 1, the participants' average rating was 6.86, while it ended at 7.611. This means the overall increase is 10.95\%. When analyzing the other weeks, weeks 4 to 5 are the only weeks with a decrease in value, from 7.310 to 7.211 (-1.37\%). Furthermore, the next biggest difference is between weeks 5 and 6, where the rating increased from 7.211 to 7.483 (3.77\%). The summary of these calculations can be seen in Table~\ref{tab:statsLHMM}.

\begin{table}[hbtp!]
\centering
\caption{Daily journal ratings}
\begin{tabular}{|l|l|l|l|}
\hline
\textbf{Low - W.1} & \textbf{High - W.8} & \textbf{Mean} & \textbf{Median} \\ \hline
6.867              & 7.6111              & 7.291         & 7.283           \\ \hline
\end{tabular}

\label{tab:statsLHMM}
\end{table}

\subsection{Thematic analysis}

\begin{table}[hbtp!]
\centering
\caption{Themes and sub-themes from thematic analysis}\label{tab:themes}
\label{tab:themes2}{\small 
\begin{tabular}{|p{4cm}|p{4cm}|}
\hline
\textbf{Theme} & \textbf{Sub-Theme(s)} \\\hline \hline
Temporal Dynamics in Self-Management: An Exploration through Reflection and Action & Struggles with Stress and Discomfort \newline Positive Transformation \\ \hline 
Self-Enhancement, Resilience, and Engagement & Mental resilience\newline Physical action\\ \hline
Workplace Experiences and Emotional Expressions & Positive: Efficiency\newline Negative: Inefficiency\\ \hline
\end{tabular}
}
\end{table}

\subsubsection*{\textbf{Theme 1: Temporal Dynamics in Self-Management: An Exploration through Reflection and Action}}

This theme captures the journey of participants through two distinct phases of the study. The first half portrays struggles with stress and discomfort, marked by difficulties in handling negative emotions and implementing positive changes. The second half, however, reveals a noticeable shift towards self-reflection, proactive habits, and prioritisation of mental well-being. The overall transformation in participants' well-being is highlighted through their evolving ability to reflect on and address their emotional states, leading to positive changes in habits and overall mental health.

\hspace*{0.4cm}\textit{Sub-theme: Struggles with Stress and Discomfort (First half of the study week 1-4):}

This theme captures instances where individuals reported lower levels of general well-being, reflecting a struggle to cope with stress and discomfort. The quotes illustrate participants' difficulties, such as exhibiting avoidant behaviours, feeling overwhelmed with work, and experiencing unproductive days. Despite engaging in self-reflection, participants found it challenging to implement effective strategies to alleviate their stress, leading to an overall impact on their mental health. The theme emphasises the early hurdles participants faced in managing their well-being during the initial weeks of the study.

\interviewquote{I had more avoidant behaviours than of late. Still not opened online shopping packages delivered 2 days ago}{P10}

\hspace*{0.4cm}This quote is a good, general representation of days rated low during the first half (week 1 -week 4) of the study. The participant demonstrates some self-reflection in their behaviour and general mood but is still stuck in ther mindset. The second sentence further highlights the sub-theme avoidance, and how they have struggled with opening their arrived package. For the general participant, being able to self-reflect further than their feeling of discomfort or that something is wrong is rare early on in the study.

\interviewquote{...I really don't feel like working. I don't even know what to do. I do have an endless to do list. I just don't feel like doing absolutely anything on it.}{P7}

The entry above is from week 3. P7 describes being overloaded with work which results in them feeling unproductive. P7 reflects over being unproductive, but still, just as the earlier submission, struggles to come up with an action or solution.

\interviewquote{I woke up slightly upset, I can’t explain why I was so frustrated. When started working I got distracted and forgot about it for a while, then the day was almost over and I had some time free in the working calendar, but that time only got me terribly distracted. I only wanted that feeling to end, I tried to meditate, to read, I got a shower to make my thoughts and feelings go away. Anything I tried work, I was simply and deeply mad. In the afternoon, after the working day was fully over I decided to sleep, was the only thing that more or less worked. Very strange and somehow uncontrollable sensation.}{P3}

\hspace*{0.4cm}The excerpt above is a journal entry from week 3. The entry implies P3 is experiencing discomfort but cannot explain why. The discomfort leads to an unproductive day at work, further frustrating P3. What differs from the previous journal entries described is that P3 tried to take action to alleviate the stress but without any success. The entry is still from the first half of the study, and therefore, it makes sense that P3 was trying different methods for relieving themselves of the feeling, even though the attempts failed. This quote can be likened to the quote by P10, as they both showcase self-reflection over the feeling of discomfort, further highlighting the general perception of the participants' stress and discomfort.\\
\hspace*{0.4cm}The preceding three examples highlighted the challenges participants faced in managing stress and discomfort during the initial phase of the study. More often than not, individuals attempting to address these issues encounter difficulties. Participants generally struggled to make significant changes to their habits that would positively influence their mental well-being. Although a few participants succeeded in this endeavour, it was not the prevailing trend among the general cohort.

\hspace*{0.4cm}\textit{Sub-theme: Positive Transformation (Second Half: week 5-8):}
During the second half of the study, the general participant managed to change their habits to positively impact their mental well-being. Improvements could be seen in how participants self-reflected and also how a lot of them began prioritising their mental well-being.

\interviewquote{Even though I got pressured at work by many people, I remained calm at work and just fixed what needed fixing. I am proud that I have managed that situation well. Back in December, I would have lost my need as I felt rather burnt out back then. I have been actively working on managing stress since October last year as since then, my workload has tripled due to reorganization.}{P12}

\hspace*{0.4cm}P12 reflects on a challenging work situation where they experienced pressure from multiple sources. Despite the stress, P12 highlights their ability to remain calm and effectively address the issues. Their sense of pride in managing the situation contrasts with their experience in December when they felt burnt out and less capable of handling such pressures. The participant attributes their improved stress management to active efforts initiated since October of the previous year, coinciding with a significant increase in workload due to organizational changes. This narrative suggests a positive transformation in P1's stress-coping mechanisms and resilience.

\interviewquote{What a day! Sun was up and I took a long lunch walk. It was amazing! I feel so much better after the walk and meetings went smoother}{P8}

\hspace*{0.4cm}P8 went out for a lunch walk in the sun and got very energised; this led to the participant feeling better in their meetings afterwards. To take a longer break with some physical activity seemed to be the key to this day, the day was rated at a max score of 10. Indicating that even meetings can go smoothly when properly caring for their body.

\interviewquote{I keep working on my journal, on getting some energy back from topics that stayed open, but required a finalization. Meditation has helped me greatly, I have been able to focus more on the present moment, have more focus and feel genuine joy more often...}{P3}

\hspace*{0.4cm}In the provided quote, there is an elucidation of a transformation in P3 cognitive self-perception. This participant describes an augmented experience of genuine joy and an improved ability to remain present in the moment, resulting in positive effects. Consequently, this particular case is noteworthy as it exemplifies an individual with an established capacity for reflective practices who experienced further personal growth during the study, albeit not without encountering challenges.

\interviewquote{I was very pleased to see today how I handled unexpected events at work. I calmly reacted to a couple of emails today that in the past would have triggered my anxiety. I think what worked was having a mindset of focusing on what is important for me to do today without fixating on distant tomorrow's.}{P11}

\hspace*{0.4cm}The quote above shows how P11 expresses satisfaction with their ability to navigate unforeseen challenges at work. The participant notes explicitly their calm reaction to emails that would have previously triggered anxiety. The key to their improved response is attributed to adopting a mindset centered on the immediate tasks at hand, without becoming overly preoccupied with future uncertainties. This mindset shift reflects a positive change in P11's approach to managing unexpected work situations, indicating increased resilience and a more focused perspective on daily priorities.

Generally, participants showcased a better ability to self-reflect later on in the study. Furthermore, participants engaged in more and more physical self-care as the study prolonged. 

\subsubsection*{\textbf{Theme 2: Self-Enhancement, Resilience, and Engagement}}

The theme describes the actions the participant took to increase their well-being. This was, in general, done through meditation, exercises, and breathing practises that were introduced in the study. The theme not only captures physical activities but also explores the mental strategies employed by participants to navigate their internal struggles, reflecting a holistic approach to self-governance. In essence, this theme gathers indications of action taken for the sake of bettering oneself (self-care). \\
\hspace*{0.4cm}\textit{Sub-theme: Mental Resilience:}
Under this sub-theme, participants showcased mindful reflections- thoughts that led them to insights about their emotions and reactions and what that means to them. An example of this would reflections by P13 in the quote below.

\interviewquote{I am more conscious and aware not to stress myself during work. Even tough, today I had lunch only 14pm because I wanted to finish several small tasks in the morning. I must admit I was a bit stressed to squeeze too much. I am trying to breathe deeply without holding air in the middle. The first session of the study was a good reminder that I have the tools and knowledge to increase my inner well-being.}{P13}

In the quote above, the participant acknowledges an increased awareness of stress management at work. Despite experiencing stress while trying to accomplish tasks, they note a conscious effort to practice deep breathing. The mention of the first study session serves as a reminder to possess the tools and knowledge to enhance inner well-being. Overall, the participant recognises the positive impact of the study intervention in empowering them to navigate workplace stress more effectively.

\interviewquote{I was stressed today and a little bit down, so to overcome with the situation I try to think what I am grateful while I am walking, and it helps me to same degree.}{P14}

Participants shared strategies for coping with stress and a slightly low mood. One example is P14's journal entry: When faced with these emotions, they practice thinking about what they are grateful for while walking. This intentional focus on gratitude during the activity serves as a helpful mechanism, providing some degree of relief from the stress and negative emotions they are experiencing.

\interviewquote{Felt really stressed about the tasks and that I am not progressing fast enough. The presentation about time management and seeing that others also struggle with it helped, followed by the breathing practice that was a strange experience at first, but breathing, doing nothing and listening to the music certainly helped to disengage from work.}{P1}

Performing the breathing practises helped P1 distance themselves from work to relax and energise. Seeing this was a stressful day, the participant reflected on how the session helped them feel better. In the long run, this reflection leads to the participant being more aware, which is the first step in taking action. Another interesting take from this quote is that the daily score was 7/10, which is high. This proves that taking action will provide a better daily perception.

\hspace*{0.4cm}\textit{Sub-theme: Physical action:}
The sub-theme focuses on instances where participants actively engaged in physical activities as a form of self-care. It encompasses intentional movements and exercises aimed at promoting well-being and managing stress.

\interviewquote{...The meditation practice helped energise me a bit. I’m grateful to have started my exercise routine again}{P4}

The example from P4's journal entry highlights how engaging in meditation and restarting an exercise routine energises the participant. This participant achieved to incorporate two practices that interconnect mental and physical well-being in their self-care practices.

\interviewquote{Running in the forest without disturbance from humans or human noises makes me feel ``empty" in my mind and more present. Just me and my dog to bother about.}{P15}

In the quote above, the engagement in physical activity, specifically running, is a crucial element in the participant's experience. The significance lies in the positive impact that this physical action has on the participant's mental state. The use of the term `empty' in the context of the participant's mind suggests a state of mental clarity, free from clutter or distractions. The participant finds a sense of presence and focus, emphasising the solitude and tranquillity of the natural environment.

\subsubsection*{\textbf{Theme 3: Workplace Experiences and Emotional Expressions}}
This theme explores participants' emotional responses and attitudes toward their work life, aiming to uncover the subjective experiences associated with their professional roles. The objective is to gain insight into how individuals engage with and interpret their work environments on an emotional level. Two sub-themes ``Triumphs in Productivity" and ``Navigating Workplace Challenges" emerged that portrait different work experiences. 

\hspace*{0.4cm}\textit{Sub-theme: Triumphs in Productivity:}
This sub-theme encapsulates instances where participants positively acknowledge task completion, expressing pride and satisfaction in their achievements. Additionally, it examines participants' adeptness in prioritising their daily activities. An illustrative example involves a participant who deliberately opted for a lunchtime walk instead of persisting in a work task, thereby seeking revitalisation and intending to enhance subsequent task efficiency.  

\interviewquote{Manage to do both fruitful work and physical activity even though I anticipated not to be able to find time for both}{P5}

Contrary to initial expectations, P5 expresses a triumph in productivity by successfully accomplishing both meaningful work and engaging in physical activity. This quote reflects effective time management and balancing work and self-care.

\interviewquote{Today was a good start to the week. I kept my energy up by wiggling whenever I got up from my desk, and  even if i didn't accomplish my exact goals, I was satisfied with the work I got done.}{P16}

P16 acknowledges a positive start to the week, emphasising energy maintenance through physical movement. Despite not achieving specific goals, there is satisfaction in the completed work. The quote indicates a focus on well-being and a flexible, adaptive approach to productivity.

\interviewquote{I did the most important tasks (rocks) I had set for the day which made me feel great, so much better than the overwhelming feeling I often have when looking through my six page to do list. Also I did some additional things which was great}{P17}

The quote above shows how P17 reflects on completing crucial tasks for the day (``rocks"\footnote{This refers to the Rock, Pebbles, and Sand Analogy for Time Management, source unknown. For details, see e.g.~\hyperref[https://www.thebusyllama.com/rocks-pebbles-sand/]{https://www.thebusyllama.com/rocks-pebbles-sand/}}), resulting in a positive emotional response. The accomplishment contrasts with the overwhelming feeling often associated with a comprehensive to-do list. Additionally, completing extra tasks adds to the overall sense of achievement.

These quotes collectively highlight the participants' ability to navigate and triumph over challenges, finding satisfaction in their productivity and well-being.

\textit{Sub-theme: Navigating Workplace Challenges:}
Throughout the study, participants described stressful working situations. During the first half, it was more evident, as the general participants struggled to cope with work challenges. Below, a few examples of this will be shared to give an overview of the general participants' struggles and the difficulties coping with them.

\interviewquote{Today just felt exhausted and defeated. All the tiredness building up since Friday turned to a sludge today. I felt down idly watching wedding videos and went down a mental slide feeling like none of my friendships are very deep anymore. I kept getting distracted during my attempts to work or write, and made the same mistake 4 times in a row talking to someone. It didn't feel like I was on autopilot, it just felt a giant blah.}{P6}

In the excerpt above, P6 describes feeling low on energy, which, results in avoiding their working tasks. P6 describes the result of being low on energy, but does not reflect on the fact that the stress caused a ``mental slide". When in this mind-state P6 describes not being able to accomplish anything work-related, which means they are able to self-reflect, but not enough to find a solution to their problem.

\interviewquote{I did not get that much done today... even though I met a lot of people. So it felt like time went really quickly today and I did not achieve much even though I have been at work physically.}{P9}

 P9 did not feel present and was inefficient at work. They acknowledge that they were physically present but, their mind was elsewhere. Compared to P6, P9 displays also self-reflection, but both participants' self-reflections do not provide a solution.

\section{Discussion}\label{sec:d}
This section depends on the study's outcomes. It addresses the three research questions. Firstly, we explore how the intervention influenced participants' well-being (RQ1).

The average daily rating for all participants was increased from 6.86 (week 1) to 7.61 (week 8). This meant that the overall increase was 10.95\%. The thematic analysis also yielded positive results. It indicated that the tools provided in the study were useful and helped the participants perceived productivity to increase.

The results can be compared with studies from other fields as well as within IT. 
Outside of IT, Janssen et al.~\cite{janssen2018mindfulness} performed a systematic review of the effects of a clinical mindfulness program, which saw a decrease in the participants' emotional exhaustion, stress, psychological distress, depression, anxiety, and occupational stress. We see the participants perceiving themselves as less stressed over the course of the program.

Within IT, our results confirm what Bernardez et al.~\cite{bernardez2020effects,bernardez2023impact} find, namely increased well-being and hence potentially more resilience towards stress in the long run.

\cristysbox{\textbf{Take away message from RQ1:} The participants' mental well-being increased as the study progressed. The result indicates that the intervention on a general level, lead to a positive change in the participants' well-being.}


Furthermore, participants demonstrated an enhancement in perceived productivity (RQ2), corroborating findings from experiments on mindfulness, such as the results by Bernárdez et al.'s~\cite{bernardez_experimental_2018} from two experiments to evaluate mindfulness, which also concluded a positive impact. 

By combining the results of RQ1 and RQ2, we see participants' perceived productivity increased as stress and anxiety decreased. Akula \& Cusick~\cite{akula_impact_2008} found a similar result and suggested it may impact overall software quality. We also confirm Maudgalya et al.~\cite{maudgalya_workplace_2006} and Fucci et al.~\cite{fucci_need_2020}, who all point to stress as a reason for a decrease in productivity. \\

\cristysbox{\textbf{\textbf{Take away message from RQ2:}}
General participants' perceived productivity has increased from the first to second half of the study. Participants learned to incorporate different tools to help them stay efficient throughout the day.}


Regarding the changes in participants' daily perceptions (RQ3), the results indicated the overall daily perception progressed for the better. The participants benefited from the study by being able to reflect more positively, which increased their well-being. Initially, negative aspects were mentioned without reflective insights in the first half, while in the latter half, participants began recognising opportunities for self-improvement through habit changes. This shift in reflection suggests a positive evolution in participants' daily perceptions throughout the study.

Something we found common as a whole was that the participants focused more on the negative aspects of the day rather than the positive. It is hard to eliminate all the negativity from your life but seeing the positive gave us the results that the day gets better. Leary describes the paradox of becoming too self-reflective, which can result in a person being very egocentric, as well as more stressed~\cite{leary2007curse}. This extreme would also be known as ``rumination''. Rumination is repetitive, prolonged, and recurrent negative thinking about one's self, feelings, personal concerns and upsetting experiences~\cite{watkins2020reflecting}. It usually happens more in purely mental practices and not in embodiment practices (like the breathing practice in the study at hand) where the focus lies in staying highly perceptive of what is going on in the body in terms of physical sensations and emotions. 
Nevertheless, when becoming more self-aware, one might analyze deeper regarding all feelings which could bring stress to the table. To mitigate this, we repeatedly remind participants to focus on the physical sensations caused by a perceived emotion instead of following a train of thoughts. \\

\cristysbox{\textbf{\textbf{Take away message from RQ3:}}
The participants' overall daily perception progressed positively over time, with increased self-reflection during the second half of the study.}

\hspace*{0.4cm}\emph{Impact on Software Engineering.}
Previously discussed studies~\cite{bernardez_controlled_2014,amin_software_2011} mention how the software industry is a stressful environment to work within. When discussing stress in the workplace, it tends to be correlated with efficiency or productivity~\cite{kompier1999preventing}. 
We argue the impact goes way beyond traditional productivity measurements and should include employee retention and sick time.

Notably, this is in the first series of studies using breathwork as an intervention in IT and amongst the first empirical studies on breathwork in any field. It offers a novel approach to increase well-being within the IT context. This research also contributes to limited literature on mindfulness interventions and paves the way for exploring and applying mindfulness-based strategies in IT. It extends the current understanding and sets the stage for future studies to investigate diverse approaches for enhancing the well-being and, as a second-order effect, the retention and the productivity of IT workers.

\emph{Threats to validity.}
In terms of external validity, the sample size is too small to generalize the results. During week 2 of the study, there were a total of 111 journal entries. During week 8, that number dropped to 36. 
In regard to internal validity, there is a possibility for selection bias when presenting excerpts of the qualitative data. We mitigated this by defining a rigorous data analysis method: All data was coded individually and then compared and discussed the results with each other. Furthermore,  breathing practices were the largest part of the study but not the only. However, the journal data contained a lot of positive feedback for the breathing practices. It is possible that participants who did not perceive benefits dropped out without leaving explicit comments on that in their journals. That said, we received notifications from many dropouts that didn't have enough time for the study.
With regard to construct validity, being stressed and having a decreased level of well-being can be more palpable because of the ongoing global pandemic (COVID-19)~\cite{qiu2020nationwide}.
This might skew the data in a way that participants feel more stressed compared to “ordinary” living circumstances.

\section{Conclusion and Future Work}\label{sec:c}

This paper presents the analysis of the journal entries of a study that investigates mindfulness as a stress-alleviating procedure for workers within the IT sphere. The data consisted of 610 journal entries by 65 participants. The journal entries were then analysed using thematic analysis to answer the research questions.

The general result indicates that breathing practices in combination with reflective journaling have a positive impact on IT workers' well-being and productivity, but further research is needed to establish it as a fact. If more research reaches the same conclusion, mindfulness could serve a real purpose for the IT industry as a whole, as increasing productivity and mental well-being is of interest to both employer and employee.

\emph{Future Work.}
Since this study was organized during a global pandemic and done in a remote setting, it would be very interesting to conduct a similar study in-house at an IT company. Two of the authors are currently conducting an in-house study evaluating a weekly yoga program with a software development company.

\bibliographystyle{ieeetr}
\bibliography{thesisref.bib}

\end{document}